\documentclass{mem}
\usepackage{natbib}\usepackage{txfonts}\usepackage{balance}
\usepackage{graphicx}
\usepackage[a4paper]{hyperref}
\idline{75}{282}
\begin{document}
\def\teff{$T\rm_{eff }$}
\def\kms{$\mathrm {km s}^{-1}$}

\title{A comprehensive survey of variable stars in the globular cluster NGC 362}

   \subtitle{}

\author{P. \,Sz\'ekely\inst{1}, L. \,L. \,Kiss\inst{2,3}, B. \,Cs\'ak\inst{1,4}, A. \,Derekas\inst{3,5}, T. \,R. \,Bedding\inst{3} \and K. \,Szatm\'ary\inst{1}}

  \offprints{P. Sz\'ekely}

\institute{
Department of Experimental Physics and Astronomical Observatory, University of Szeged, Szeged, D\'om t\'er 9., 6720, Hungary
\email{pierre@physx.u-szeged.hu}
\and
Department of Experimental Physics
and Astronomical Observatory, University of Szeged, Szeged, D\'om t\'er 9.,
6720 Hungary (on leave)
\and
School of Physics, University of Sydney, NSW 2006, Australia
\and
Harvard-Smithsonian Center for Astrophysics (CfA), 60 Garden Str., Cambridge, MA 02138, USA
\and
School of Physics, Department of Astrophysics and Optics, University of New South Wales, Sydney, NSW 2052, Australia
}

\authorrunning{Sz\'ekely }

\titlerunning{Variable stars in NGC 362}

\abstract{
We present the first results of our variability survey for the globular
cluster NGC 362. We found numerous variable stars in the field including 23 RR Lyr stars, 4 eclipsing binaries and 5 other pulsating stars. Four of the RR Lyrae stars are located behind the cluster, possibly in the farthest extensions of the SMC.

}
\maketitle{}

\section{Introduction}
Although globular clusters are primary testbeds of modern astrophysics due to large number of stars with the same age and composition, there are still a number of unstudied clusters, mostly in the southern hemisphere. We have been carrying out a CCD photometric survey project of southern globular
clusters since mid-2003. Here we present the first results for NGC 362, which is one of the brightest unstudied southern clusters, located in front of the outer edge of the Small Magellanic Cloud (SMC).

\section{Observations and data analysis}
Our V-filtered photometric observations were carried out between July 2003 and October 2004 on 17 nights at Siding Spring Observatory, Australia.
We used the 1m ANU telescope equipped with the Wide Field Imager. In addition to the photometry, we measured radial velocities for the brightest RR Lyrae stars from spectra taken with the 2.3m ANU telescope.

Instrumental magnitudes were obtained via a semi-automatic pipeline using daophot tasks for PSF photometry. Periods for variables were determined with a combination of Fourier analysis, phase dispersion minimization \citep{stellingwerf1978} and string-length minimization \citep{lafler1965}.

\section{Discussion}

Our results can be summarized as follows.

\begin{itemize}

\item From the ensemble photometry of over 10,000 stars, we found 19 RRab/RRc variable stars (of which 16 are new discoveries) with very similar mean apparent
brightnesses, so that they are all members of the cluster. The measured radial velocities for 5 RR Lyraes supported this conclusion for having the same $\sim$200 km/s mean velocity as the  cluster itself \citep{fischer1993}.

\item A large fraction of RR Lyrae stars show periodic amplitude and/or phase
modulations, the so-called Blazhko-effect, which is still one of the greatest
mysteries in classical pulsating stars (e.g. \citealt{chadid2004}). In our sample, two variables, V6 and V18 (shown right), have the strongest light curve modulations, while further five stars (V9, V12, V13, V19 and V31) exhibited subtle changes during the 15 months of observations.

\item We found a number of variable stars belonging to other classes. Interestingly, we also identified a few RR Lyrae stars that are several
magnitudes fainter than those in the cluster. We suspect these to belong to the outher halo of the Small Magellanic Cloud; a more secure membership
investigation is in progress.

\item We found 5 long period ($>$1 d) RR Lyr-like variables, which are either candidate above-horizontal-branch stars (AHB stars, \citet{diethelm1990}) or short-period Type II Cepheids.

\item We also discovered several short-period eclipsing binaries, most likely in the galactic foreground of the cluster.

\item A number of stars on the red giant branch showed evidence for variability on a time-scale of 10-50 days, suggesting the presence of RGB pulsations predicted by \citet{kiss2003}.

\end{itemize}

\begin{figure}[t!]
\resizebox{\hsize}{!}{\includegraphics[clip=true]{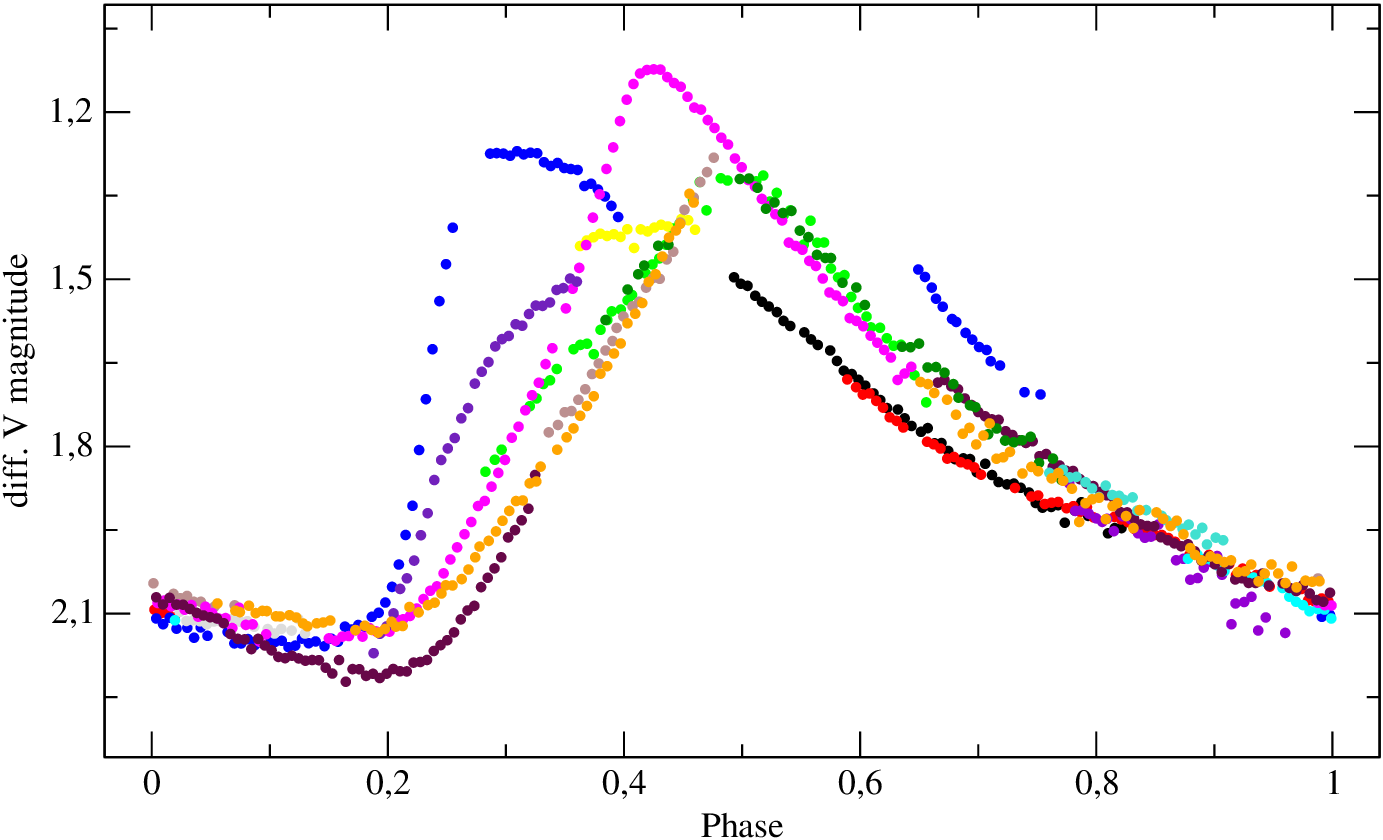}}
\caption{\footnotesize The phase diagram of V6.}
\label{v6}
\vskip2mm
\resizebox{\hsize}{!}{\includegraphics[clip=true]{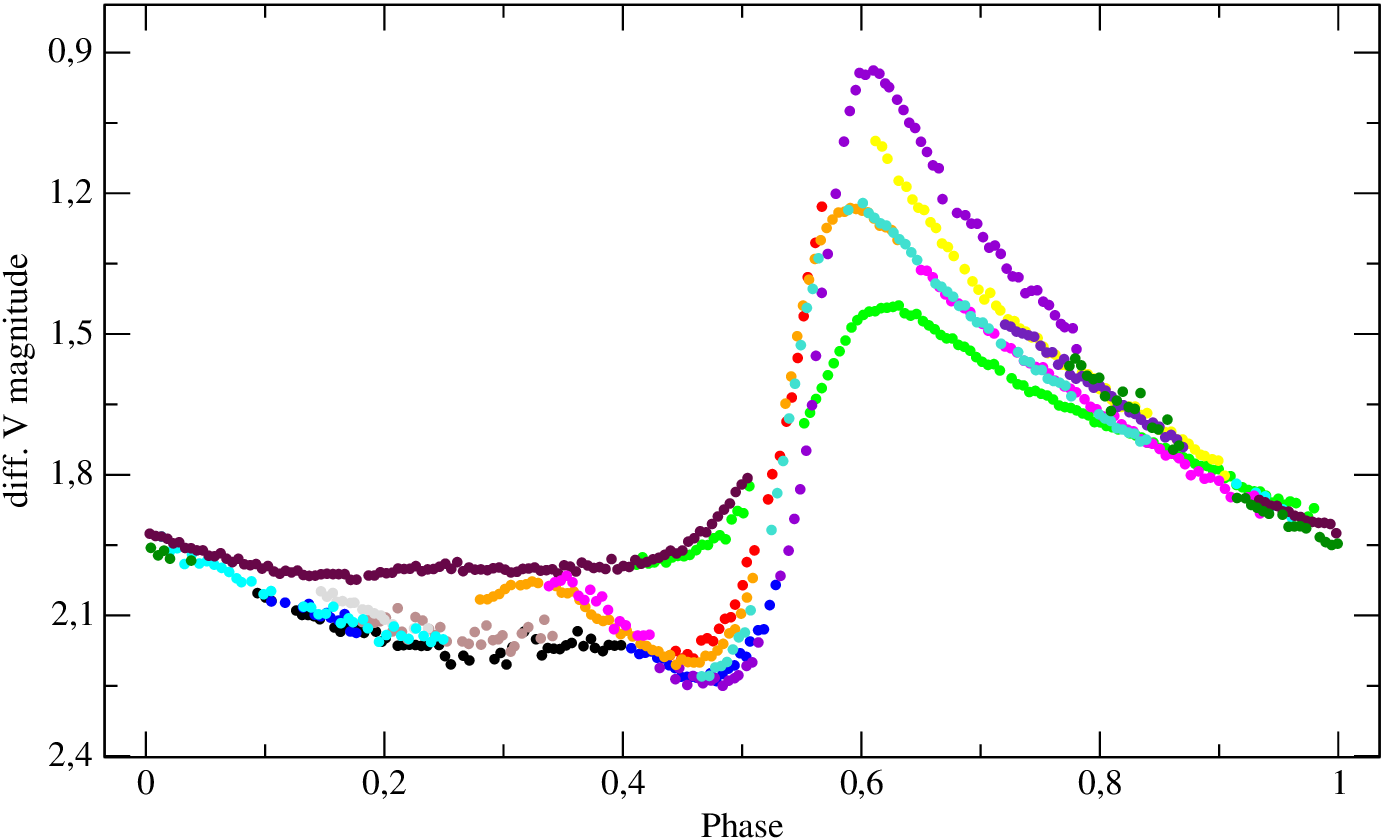}}
\caption{\footnotesize The phase diagram of V18.}
\label{v18}
\end{figure}

\begin{acknowledgements}
Our work was supported by the Australian Research Council and the Hungarian OTKA Grant \#T042509. LLK is supported by a University of Sydney Postdoctoral Research Fellowship. The NASA ADS Abstract Service was used to access data and references.
\end{acknowledgements}

\bibliographystyle{aa}

\end{document}